# Optimal Layered Defense For Site Protection

Tsvetan Asamov[1], Emre Yamangil[1], Endre Boros[1], Paul Kantor[1,2,3,4,5,6], and Fred Roberts[5,6]

[1]Rutgers Center For Operations Research
[2]Rutgers School of Communication and Information
[3]Department of Systems and Industrial Engineering, University of Wisconsin, Madison
[4]Paul B Kantor, Consultant
[5]DIMACS (Center For Discrete Mathematics and Theoretical Computer Science)
[6]CCICADA (Command, Control, and Interoperability Center for Advanced Data Analysis)

**Abstract**

We present a model for layered security with applications to the protection of sites such as stadiums or large gathering places. We formulate the problem as one of maximizing the capture of illegal contraband. The objective function is indefinite and only limited information can be gained when the problem is solved by standard convex optimization methods. In order to solve the model, we develop a dynamic programming approach, and study its convergence properties. Additionally, we formulate a version of the problem aimed at addressing intelligent adversaries who can adjust their direction of attack as they observe changes in the site security. Furthermore, we also develop a method for the solution of the latter model. Finally, we perform computational experiments to demostrate the use of our methods.

# Acknowledgement

This material is based upon work supported by the U.S. Department of Homeland Security under Grant Award Numbers 22STESE000010101 from DHS Office of University Programs via Northeastern University, DHS 2009-ST-061-CCI002 from DHS Office of University Programs to Rutgers University, and HSHQDC-13-X-00069 from DNDO.

Disclaimer: The views and conclusions contained in this document are those of the authors and should not be interpreted as necessarily representing the official policies, either expressed or implied, of the U.S. Department of Homeland Security.

# 1   Introduction

We study the problem of defending a target such as a stadium or a large gathering place with multiple access paths. In practice, the notion of "layered defense" is commonly used to describe the idea that we have an outer perimeter where we first seek to capture dangerous entities (vehicles, people, cargo), then perhaps a middle perimeter or perimeters where we do the same thing using different methods and perhaps information gathered from the outer perimeter, and then an inner perimeter where we again use different methods and information gathered from earlier perimeter defense. Thus, as vehicles approach a stadium, we might do license plate reading; in a middle layer or layers we use radiation detectors or behavioral detection of patrons after they have parked their cars; then an inner layer we use metal detection through wanding or walkthrough magnetometers or where we inspect bags or pat-down patrons. We seek to make this idea of layered defense precise in an abstract, simplified way.

We speak abstractly of "sensors" at each layer of defense, but understand that our "sensors" could be physical sensors but also tests of different kinds such as behavioral observation. Our approach is based in an increasing literature that deals with inspection processes using a number of potential tests, for example at ports of entry. In the past few years numerous techniques for sensor optimization of port-of-entry inspection have been explored in the literature [6, 8, 7, 14, 17, 12, 1, 5, 9, 16, 13]. Several authors have reported numerical results that demonstrate significant improvement over straightforward inspection approaches [6, 7, 12, 1, 16, 13]. In line with existing practices, most researchers have assumed that the vast majority of the inspected items and people are perfectly legal and only a very small proportion of the incoming flow is harmful. Under such circumstances, the sensor operating cost (though not the capital cost) is usually only a small fraction of the overall cost of the inspection operation. The bulk of the total cost and time spent is attributed to a thorough inspection procedure that is performed on potential suspicious items and individuals. Such a situation is usually encountered in airport security checkpoints, border crossings, maritime port inspection stations, large sports stadiums, etc.

In mathematical terms, the problem of layered site security is quite different from optimizing a set of sensors searching for illegal cargo at a port-of-entry. Unlike much of the existing inspection optimization work which considers two distinct populations of inspected items, i.e. legal and illegal, in our model we only consider the latter. We work under the assumption that we can incorporate the processing cost of occasional encounters of legal traffic into the overall cost curve for detecting contraband.

## 2 Mathematical Model

To develop our ideas, we have formulated a model of a perimeter defense of the target with two layers of defense where we have a limited budget for surveillance and we need to decide how much to invest in each layer and where to invest it if there are several locations where we might do inspection in each layer. Defense at the outer layers might be less successful but could provide useful information to selectively refine and adapt strategies at inner layers. Arranging defense in layers so that decisions can be made sequentially might significantly reduce costs and increase chance of success. Monitoring at an outer layer could not only hinder an attacker but could provide information about the current state of threat that could be used to refine and adapt strategies at inner layers. There is a complex tradeoff between maximizing the cost-effectiveness of each layer and overall benefits from devoting some efforts at the outer layer to gathering as much information as possible to maximize effectiveness of the inner layer.

To give a stylized abstract version of what we have in mind, consider Figure 1, where we show a target in the middle, threats arrive via four inner channels and each reachable from two outer flows of vehicles, patrons, etc.

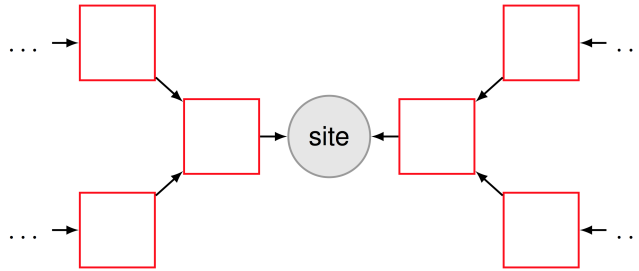

**Figure 1:** An abstract model of layered defense showing a target in the middle, threats arriving via two inner channels, and each reachable from two outer flows of vehicles, patrons, etc. We concentrate in this paper on the problem with two layers of defense, where each security layer has a number of sensors placed on possible paths of incoming illegal flow of vehicles and/or patrons. Inner layers are composed of sensors that are used to detect units that have managed to infiltrate outer layers undetected. Every interior sensor is connected to one or more sensors in the immediately preceding outer layer in the sense that it is responsible for backing up those sensors, i.e., the goal of the interior sensor is to discover traffic which has remained undetected by those outer sensors. In order for an illegal unit to penetrate the system, it would need to remain unnoticed at all layers of inspection. We denote the set of sensors in the internal layer of defense with $I$, and the set of sensors in the external layer with $J$. We assume that sensors in the set $I$ share a limited total resource budget $X$ and sensors in the set $J$ share a limited total budget $Y$. More subtle models allow one to make decisions about how much budget to allocate between the inside and outside layers. Our objective is to develop optimization methods to determine the optimal allocation of resources to security sensors in such a manner that the expected detection rate of incoming threats is maximized. For modeling purposes we employ the following assumptions:

- There exists only one type of violation that we are protecting against.
- The expected number of contraband units on each incoming path is a known parameter. For the outermost perimeter sensor $j$, we denote the incoming contraband flow with $F_j$.
- For each sensor $i \in I$, located at the inside security layer, we know the function $D_i^x(x)$, which specifies the detection rate at the sensor for contraband items if the total amount of resources made available to the sensor is $x$, and similarly for each $j \in J$ we know the detection function $D_j^y(y)$. In this paper we assume that the detection functions are specified as concave increasing piecewise linear functions. Thus, we do not require the detection functions to be differentiable everywhere, which is an important property of our method. We also assume that the resources are normalized to take values between 0 and 1.
- All sensors at a given layer share a limited common resource. For example, an outside perimeter could be supported by a fixed number of infrared motion detectors or license plate readers, while an inside perimeter could consist of walkthrough magnetometer tests or security guards conducting wanding of patrons.

Our goal is to allocate the total outside resources among individual sensors and allocate the total inside resources among individual sensors in order to maximize the detected illegal flow. Thus we arrive at the following mathematical formulation:

$$
\begin{aligned}
\max_{\mathbf{x},\mathbf{y}} \sum_{i \in I} &\left\{ \left( \sum_{j \in N(i)} F_j \cdot D_j^y(y_j) \right) + D_i^x(x_i) \left( \sum_{j \in N(i)} F_j (1 - D_j^y(y_j)) \right) \right\} \\
\text{s.t.} &\sum_{i \in I} x_i \leq X \\
&\sum_{j \in J} y_j \leq Y \\
&x_i \geq 0, \forall i \in I \\
&y_j \geq 0, \forall j \in J
\end{aligned}
\tag{1}
$$

where $N(i)$ denotes the set of outside sensors adjacent to inside sensor $i$. Here, the first sum over the outside neighbors $j$ of $i$ gives the flow that is captured at $j$ and the second sum gives the flow that is not captured at $j$ but is captured at $i$.

Now, let us examine the given objective function. Clearly, it contains mixed nonlinear product terms of detection probabilities. Moreover, since there are no pure quadratic terms, we know that in general the objective function is neither convex, nor concave. We illustraste this in Example 2.1.

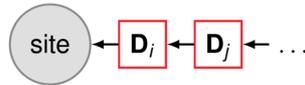

**Figure 3:** A model of layered defense showing a target with a single exterior sensor preceding a single interior sensor.

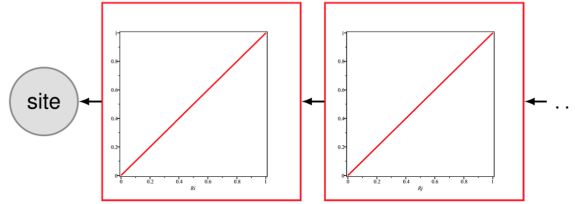

**Figure 4:** Linear detection rates at both the exterior and interior sensors of Figure 3.

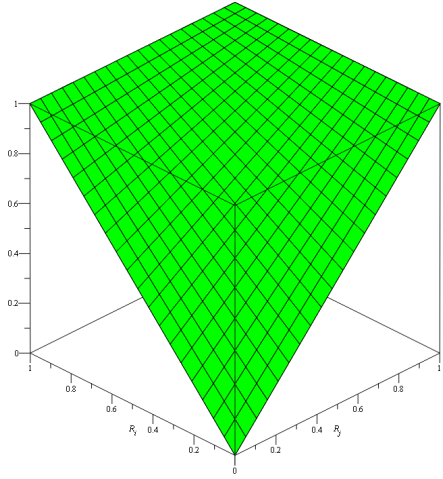

**Figure 5:** A plot of the objective function of (2) for the case of $D_i^x(x_i) = x_i$, $D_j^y(y_j) = y_j$ and $F_j = 1$. In this case three of the corners of the feasible region are optimal solutions.

*Example* 2.1 (Indefinite Objective Function). Suppose we have a single exterior sensor $j$ preceding a single interior sensor $i$, as shown in Figure 3. In this case, we would need to solve the following problem.

$$\max_{\mathbf{x_i}, \mathbf{y_j}} \left\{ F_j \cdot D_j^y(y_j) + D_i^x(x_i) \cdot (F_j \cdot (1 - D_j^y(y_j))) \right\}$$
$$\text{s.t. } 0 \leq x_i \leq X \qquad (2)$$
$$0 \leq y_j \leq Y$$

Let us for a moment consider what would happen if $D_i^x(x) = x$ and $D_j^y(y) = y$ as shown in Figure 4. In that case, $D_i^x$ and $D_j^y$ are differentiable everywhere. Thus, if we denote the objective function in problem (2) as

$$f^{i,j}(x_i, y_j) = F_j \cdot D_j^y(y_j) + D_i^x(x_i) \cdot (F_j \cdot (1 - D_j^y(y_j)))$$

then we know

$$\begin{aligned}
\nabla f^{i,j}(x_i, y_j) &= \begin{bmatrix} \frac{\partial f^{i,j}(x_i,y_j)}{\partial x_i} \\ \frac{\partial f^{i,j}(x_i,y_j)}{\partial y_j} \end{bmatrix} \\
&= \begin{bmatrix} F_j(1 - D_j^y(y_j)) \\ F_j(1 - D_i^x(x_i)) \end{bmatrix} \\
&= \begin{bmatrix} F_j(1 - y_j) \\ F_j(1 - x_i) \end{bmatrix} \\
&\geq \begin{bmatrix} 0 \\ 0 \end{bmatrix}
\end{aligned} \quad (3)$$

Therefore the objective fuction is increasing everywhere in the feasible region. Thus, we know that we would get an optimal solution to problem (2) by setting $x_i = X$ and $y_j = Y$. However, upon further inspection we can notice that if we attempted to solve the problem as a convex optimization problem, we would run into difficulties. The Hessian of the objective function has the following form:

$$\nabla^2 f^{i,j}(x_i, y_j) = \begin{bmatrix} 0 & -F_j \\ -F_j & 0 \end{bmatrix} \quad (4)$$

And its two eigenvalues are $\lambda_1 = F_j$ and $\lambda_2 = -F_j$. Hence we know that the Hessian matrix associated with the quadratic terms is indefinite.

The indefinitess of the Hessian presents a major obstacle to solving the problem with standard solvers for quadratic programming. In our study, we tried solving numerous instances using different methods implemented in the MATLAB optimization toolbox. While in some cases we were able to produce consistent output, none of the examined methods were able to overcome the indefiniteness of the Hessian matrix for all possible values of the input parameters. This created the need for the development of an alternative solution method for the problem.

## 3 Exhaustive Search Methods

A standard approach to such problems is a brute force approach that fixes a resource partition mesh and enumerates all possibilities. This exhaustive search approach would be to discretize the resource space for each sensor into a number of subintervals. Then we could examine every possible resource allocation scenario and among all feasible cases select the one that maximizes the objective function value. However, this method would be computationally intractable even for trivial cases. For example, suppose that we have four inside sensors, and each of them is related to exactly two outside sensors. Further, suppose we split the parameter search space of each sensor into one hundred discrete intervals. Then we would need to evaluate the objective function a total of $100^{4+8} = 10^{24}$ times, which is clearly unacceptable unless a very large cluster is used. Moreover, if we considered a slightly larger case of fifteen interior sensors, each supporting a couple of outside perimeter sensors, then the number of cases explodes to $100^{15+30} = 10^{90}$, which exceeds the current estimates for the number of atoms in the universe.

However, it is sufficient to discretize the parameter space for the interior sensors. Then, for each fixed set of values, we can find the optimal configuration of the exterior perimeter by solving a linear programming problem. If we take this approach, then the two above mentioned instances require, respectively, the solution of $10^8$ and $10^{30}$ small linear programming problems. While this is a significant improvement, we are still subjected to the curse of dimensionality as the number of sensors in the interior perimeter increases. Fortunately, we can overcome this challenge.

## 4 Dynamic Programming Method

To illustrate the idea behind our method, we consider the following basic example. Suppose we would like to solve problem (2) for the case when $D_i^x$ and $D_j^y$ are general piecewise linear functions, as illustrated in Figure 6.

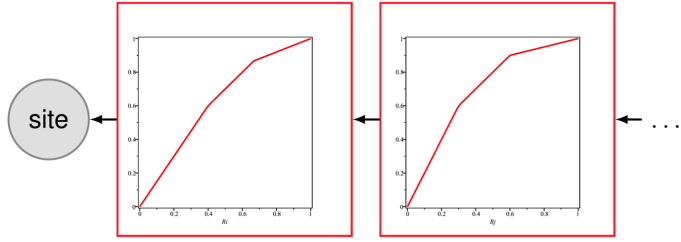

**Figure 6:** Piecewise linear detection rates at both the exterior and interior sensors of Figure 3. In that case, the objective function $f^{i,j}(x_i, y_j)$ of problem (2) is still continuous. Further, $f^{i,j}(x_i, y_j)$ is differentiable everywhere except at points correspoding to cornerpoints of the detection functions $D_i^x$ and $D_j^y$. Moreover, at points where $f^{i,j}(x_i, y_j)$ is differentiable, its gradient has the form

$$\nabla f^{i,j}(x_i, y_j) = \begin{bmatrix} \frac{\partial f(x_i, y_j)}{\partial x_i} \\ \frac{\partial f(x_i, y_j)}{\partial y_j} \end{bmatrix}$$
$$= \begin{bmatrix} c_i F_j (1 - D_j^y(y_j)) \\ c_j F_j (1 - D_i^x(x_i)) \end{bmatrix} \quad (5)$$
$$\geq \begin{bmatrix} 0 \\ 0 \end{bmatrix}$$

for some constants $c_i, c_j \geq 0$. Thus, the optimal solution is again setting $x_i = X$ and $y_j = Y$.

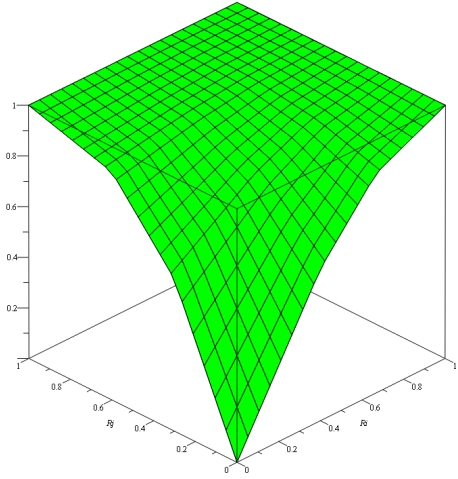

**Figure 7:** A plot of the objective function of (2) for piecewise linear functions $D_i^x(x_i)$ and $D_j^y(y_j)$.

Our solution method involves four main steps.

**Step 1:**
For a fixed $\epsilon > 0$, we create a partition $\mathcal{Y} = \{0, \epsilon, 2\epsilon, \ldots, Y\}$ of the interval $[0, Y]$, as well as a partition

$\mathcal{X} = \{0, \epsilon, 2\epsilon, \ldots, X\}$ of the interval $[0, X]$.
**Step 2:**

For every pair of sensors $i, j$ such that $i \in I$ and $j \in N(i)$, we compute $T^{i,j}(X_i, X_j)$ for each $(X_i, Y_j) \in \mathcal{X} \times \mathcal{Y}$, where we use $T^{i,j}(X_i, Y_j)$ to denote maximum amount of detected illegal contraband when inner sensor $i$ uses at most $X_i$ inner resources, and outer sensor $j$ uses at most $Y_j$ outer resources. Formally, we need to compute the optimal value of the following optimization problem:

$$T^{i,j}(X_i, Y_j) = \begin{cases} \max_{\mathbf{x_i, y_j}} \left\{ F_j \cdot D_j^y(y_j) + D_i^x(x_i) \cdot (F_j \cdot (1 - D_j^y(y_j))) \right\} \\ \text{s.t. } 0 \leq x_i \leq X_i \\ 0 \leq y_j \leq Y_j \end{cases} \quad (6)$$

We can solve an instance of problem (6) in $O(1)$ time by setting $x_i = X_i$ and $y_j = Y_j$. Thus, we can compute $T^{i,j}(X_i, Y_j)$ for each $(X_i, Y_j) \in \mathcal{X} \times \mathcal{Y}$ in $O(|\mathcal{X}||\mathcal{Y}|)$. Hence, we can plot the objective function in (2) with arbitrary precision which would ultimately allow us to solve problem (1) with arbitrary precision (see Figure 7).

**Step 3:**
Now, suppose that instead of a single outside sensor $j$, we consider two outside sensors $j_1$ and $j_2$ that are both backed up by inner sensor $i$ (see Figure 8). We use $\{j_1, j_2\}$ to denote quantities that refer to the combined system with two outside sensors $j_1$ and $j_2$. For example, $T^{i,\{j_1,j_2\}}$ denotes a table of optimal detection values for the combined system of inside sensor $i$ and two outside sensors $j_1$ and $j_2$. Further, we also use $X^{i,\{j_1,j_2\}}$ and $Y^{i,\{j_1,j_2\}}$ to denote the amount of inside and outside resources budgeted to the sensor system of inner sensor $i$, and outer sensors $j_1$ and $j_2$. Please recall that in Step 2, we computed the two tables of optimal values $T^{i,j_1}$ and $T^{i,j_2}$ (considering first the problem with only sensors $i, j_1$ and then the problem with only sensors $i, j_2$ (see Figure 9). Since they both share the same inner sensor, all we need to do in order to find the optimal detection value $T^{i,\{j_1,j_2\}}(X^{i,\{j_1,j_2\}}, Y^{i,\{j_1,j_2\}})$ for the combined system is to determine the optimal way to allocate $Y^{i,\{j_1,j_2\}}$ between sensor $j_1$ and $j_2$. Thus we have the problem of optimizing the following formulation:

$$T^{i,\{j_1,j_2\}}\left(X^{i,\{j_1,j_2\}}, Y^{i,\{j_1,j_2\}}\right) = \begin{cases} \max_{y^{i,j_1}, y^{i,j_2}} T^{i,j_1}\left(X^{i,\{j_1,j_2\}}, y^{i,j_1}\right) + T^{i,j_2}\left(X^{i,\{j_1,j_2\}}, y^{i,j_2}\right) \\ \text{s.t.} \\ y^{i,j_1} + y^{i,j_2} \leq Y^{i,\{j_1,j_2\}} \\ y^{i,j_1} \in \mathcal{Y} \\ y^{i,j_2} \in \mathcal{Y} \end{cases} \quad (7)$$

Notice that even though we consider all different values of $y_j^{i,j_1}, y_j^{i,j_2} \in \mathcal{Y}$, problem (7) can be solved in time linear in the cardinality of $\mathcal{Y}$. This is accomplished by using two index variables initialized at the two ending points of the outside resource partition $\mathcal{Y}$.

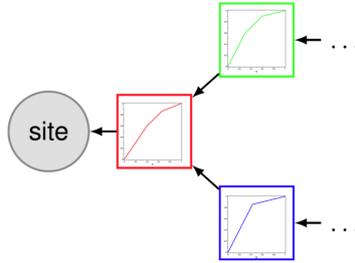

**Figure 8:** A network with two outside sensors (green and blue) and one inside sensor backing them both up.

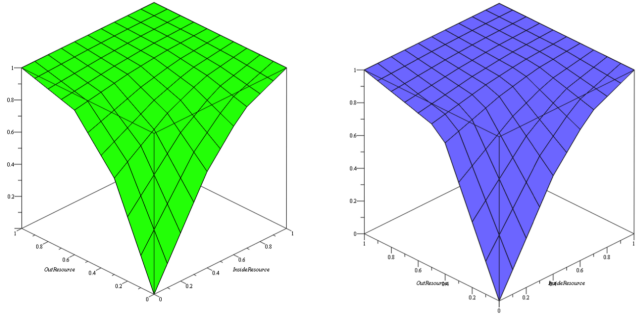

**Figure 9:** Finding separate solutions for inner sensor $i$ with each outside sensor $j_1$ and $j_2$.

If we have three outside sensors $j_1, j_2, j_3$ corresponding to inside sensor $i$, then we can find their solution matrix $T^{i,\{j_1,j_2,j_3\}}$ as follows. Once we have computed the matrix $T^{i,\{j_1,j_2\}}$ we use it as an input to equation (7), together with the matrix $T^{i,j_3}$ to generate the solution matrix $T^{i,\{j_1,j_2,j_3\}}$. By recursion, we can solve a problem instance that involves one inner sensor $i$ and any number of outside sensors. We denote with $T^i$ the solution table of optimal values corresponding to inner sensor $i$ together with all of its adjacent outside sensors $j \in N(i)$.
**Step 4:**

Suppose we are given two matrices, $T^{i_1}$ and $T^{i_2}$, that correspond respectively to two inner sensors $i_1$ and $i_2$ with their adjacent outside sensors. In order to determine the optimal detection value $T^{\{i_1,i_2\}}(X^{\{i_1,i_2\}}, Y^{\{i_1,i_2\}})$ for the combined system, we need to find the optimal way to allocate $X_i^{\{i_1,i_2\}}$ between $T^{i_1}$ and $T^{i_2}$. Thus we consider the problem of optimizing the following,

$$T^{\{i_1,i_2\}}\left(X^{\{i_1,i_2\}}, Y^{\{i_1,i_2\}}\right) = \begin{cases} \max_{x^{i_1}, x^{i_2}, y^{i_1}, y^{i_2}} T^{i_1}\left(x^{i_1}, y^{i_1}\right) + T^{i_2}\left(x^{i_2}, y^{i_2}\right) \\ \text{s.t.} \\ x^{i_1} + x^{i_2} \leq X^{\{i_1,i_2\}} \\ y^{i_1} + y^{i_2} \leq Y^{\{i_1,i_2\}} \\ x^{i_1}, x^{i_2} \in \mathcal{X} \\ y^{i_1}, y^{i_2} \in \mathcal{Y} \end{cases} \tag{8}$$

We point out that problem (8) can be solved in $O(|\mathcal{X}||\mathcal{Y}|)$ time. Finally, once we have computed $T^{\{i_1,i_2\}}$, we can proceed by recursion to solve problems involving an arbitrary number of interior and exterior sensors.

## 5 Running Time

We consider the running times of all of the four steps.
Step 1: Creating the partitions takes $O(|\mathcal{X}| + |\mathcal{Y}|)$.

Step 2: For every pair $i, j$ such that $i \in I$, $j \in N(i)$, we have to compute a matrix in $O(|\mathcal{X}||\mathcal{Y}|)$ time. Thus, step 2 takes $O(|\mathcal{X}||\mathcal{Y}||I||J|)$.

Step 3: For every $i \in I$ we perform step 3 $|N(i)|$ times, and every time we need to compute $|\mathcal{X}||\mathcal{Y}|$ number of entries, each taking $O(|\mathcal{Y}|)$. Thus the overall complexity of step 3 is $O(|\mathcal{X}||\mathcal{Y}|^2|I||J|)$.

Step 4: We need to execute this step $|I|-1$ times. Each of $|\mathcal{X}||\mathcal{Y}|$ entries in the resulting matrix takes $O(|\mathcal{X}||\mathcal{Y}|)$ time to compute. Thus, the computational complexity of step 4 is $O((|\mathcal{X}||\mathcal{Y}|)^2|I|)$.

Since the four steps are performed sequentially, we know the overall running time of the dynammic programming method is $O(|\mathcal{X}||\mathcal{Y}|^2|I|(|J|+|\mathcal{X}|))$.

# 6 Convergence

So far, we have only considered discrete approximations of the optimal detection values. In this section we show that as the partition mesh $\epsilon \to 0$, the values of the discrete approximation tables $T$ converges to the true continuous optimal detection values $t$.

Since, the discrete approximation is exact for the case of one inside and one outside sensor, we know

$$t^{i,j}(X_i, Y_j) = T^{i,j}(X_i, Y_j), \ \forall (X_i, Y_j) \in \mathcal{X} \times \mathcal{Y}$$

Consider the objective function $f^{i,j}(\cdot,\cdot)$ of the system consisting of inner sensor $i$ and its external neighbor $j$. Since $f^{i,j}$ is continuous, as well as quadratic everywhere except for a set of measure zero, we know that $f^{i,j}$ is Lipschitz continuous and we denote its Lipschitz constant with $L^{i,j}$. If we choose $L \in \mathbb{R}$ such that

$$L = \max_{\substack{i \in I \\ j \in N(i)}} L^{i,j}$$

then $L$ is a Lipschitz constant for all functions $f^{i,j}$.

Suppose $i \in I$ and $j_1, j_2 \in N(i)$, $j_1 \neq j_2$. We use $\xi^{i,\{j_1,j_2\}}$ to denote the error between the true optimal detection value $t^{i,\{j_1,j_2\}}$, and the discrete approximation $T^{i,\{j_1,j_2\}}$. We would use lower case $x^{i,j_1}$ and $x^{i,j_2}$ to denote the optimal way to split up the inside resources $X^{i,\{j_1,j_2\}}$ between $t^{i,j_1}$ and $t^{i,j_2}$, while $y^{i,j_1}$ and $y^{i,j_2}$ denote the optimal way to split up the outside resources $Y^{i,\{j_1,j_2\}}$ between $t^{i,j_1}$ and $t^{i,j_2}$.

On the other hand, we would use $X^{i,j_1}$ and $X^{i,j_2}$ to denote the optimal way to split up the inside resources $X^{i,\{j_1,j_2\}}$ between $T^{i,j_1}$ and $T^{i,j_2}$, while $Y^{i,j_1}$ and $Y^{i,j_2}$ denote the optimal way to split up the outside resources $Y^{i,\{j_1,j_2\}}$ between $T^{i,j_1}$ and $T^{i,j_2}$. Before we proceed, we need to introduce the following notation. We use $\lfloor x \rfloor$ to denote the point in $\mathcal{X}$ that is closest to $x$ from below, and we use $\lfloor y \rfloor$ to denote the point in $\mathcal{Y}$ that is closest to $y$ from below. Similarly, we use $\lceil x \rceil$ to denote the point in $\mathcal{X}$ that is closest to $x$ from above, and we use $\lceil y \rceil$ to denote the point in $\mathcal{Y}$ that is closest to $y$ from above. Then the discrete approximation error can be written as,

$$\begin{aligned}
\xi^{i,\{j_1,j_2\}}(X^{i,\{j_1,j_2\}}, Y^{i,\{j_1,j_2\}}) &= t^{i,\{j_1,j_2\}}(X^{i,\{j_1,j_2\}}, Y^{i,\{j_1,j_2\}}) - T^{i,\{j_1,j_2\}}(X^{i,\{j_1,j_2\}}, Y^{i,\{j_1,j_2\}}) \\
&= t^{i,j_1}(x^{i,j_1}, y^{i,j_1}) + t^{i,j_2}(x^{i,j_2}, y^{i,j_2}) - T^{i,j_1}(X^{i,j_1}, Y^{i,j_1}) - T^{i,j_2}(X^{i,j_2}, Y^{i,j_2}) \\
&\leq t^{i,j_1}(x^{i,j_1}, y^{i,j_1}) + t^{i,j_2}(x^{i,j_2}, y^{i,j_2}) - T^{i,j_1}(\lfloor x^{i,j_1} \rfloor, \lfloor y^{i,j_1} \rfloor) - T^{i,j_2}(\lfloor x^{i,j_2} \rfloor, \lfloor y^{i,j_2} \rfloor) \\
&= t^{i,j_1}(x^{i,j_1}, y^{i,j_1}) + t^{i,j_2}(x^{i,j_2}, y^{i,j_2}) - t^{i,j_1}(\lfloor x^{i,j_1} \rfloor, \lfloor y^{i,j_1} \rfloor) - t^{i,j_2}(\lfloor x^{i,j_2} \rfloor, \lfloor y^{i,j_2} \rfloor) \\
&= \{t^{i,j_1}(x^{i,j_1}, y^{i,j_1}) - t^{i,j_1}(\lfloor x^{i,j_1} \rfloor, \lfloor y^{i,j_1} \rfloor)\} + \{t^{i,j_2}(x^{i,j_2}, y^{i,j_2}) - t^{i,j_2}(\lfloor x^{i,j_2} \rfloor, \lfloor y^{i,j_2} \rfloor)\} \\
&\leq \sqrt{2}\epsilon L + \sqrt{2}\epsilon L \\
&= 2\sqrt{2}\epsilon L
\end{aligned} \quad (9)$$

Now, we can also bound the error in the case of three outside sensors:

$$\begin{aligned}
&\xi^{i,\{j_1,j_2,j_3\}}(X^{i,\{j_1,j_2,j_3\}}, Y^{i,\{j_1,j_2,j_3\}}) = \\
&= t^{i,\{j_1,j_2,j_3\}}(X^{i,\{j_1,j_2,j_3\}}, Y^{i,\{j_1,j_2,j_3\}}) - T^{i,\{j_1,j_2,j_3\}}(X^{i,\{j_1,j_2,j_3\}}, Y^{i,\{j_1,j_2,j_3\}}) \\
&= t^{i,\{j_1,j_2\}}(x^{i,\{j_1,j_2\}}, x^{i,\{j_1,j_2\}}) + t^{i,j_3}(x^{i,j_3}, y^{i,j_3}) - T^{i,\{j_1,j_2\}}(X^{i,\{j_1,j_2\}}, Y^{i,\{j_1,j_2\}}) - T^{i,j_3}(X^{i,j_3}, Y^{i,j_3}) \\
&\leq t^{i,\{j_1,j_2\}}(x^{i,\{j_1,j_2\}}, y^{i,\{j_1,j_2\}}) + t^{i,j_3}(x^{i,j_3}, y^{i,j_3}) - T^{i,\{j_1,j_2\}}(\lfloor x^{i,\{j_1,j_2\}}\rfloor, \lfloor y^{i,\{j_1,j_2\}}\rfloor) - T^{i,j_3}(\lfloor x^{i,j_3}\rfloor, \lfloor y^{i,j_3}\rfloor) \\
&\leq t^{i,\{j_1,j_2\}}(x^{i,\{j_1,j_2\}}, y^{i,\{j_1,j_2\}}) + t^{i,j_3}(x^{i,j_3}, y^{i,j_3}) - t^{i,\{j_1,j_2\}}(\lfloor x^{i,\{j_1,j_2\}}\rfloor, \lfloor y^{i,\{j_1,j_2\}}\rfloor) + 2\sqrt{2}\epsilon L - \\
&\quad - t^{i,j_2}(\lfloor x^{i,j_2}\rfloor, \lfloor y^{i,j_2}\rfloor) \\
&= \{t^{i,j_1}(x^{i,j_1}, y^{i,j_1}) - t^{i,j_1}(\lfloor x^{i,j_1}\rfloor, \lfloor y^{i,j_1}\rfloor)\} + \{t^{i,j_2}(x^{i,j_2}, y^{i,j_2}) - t^{i,j_2}(\lfloor x^{i,j_2}\rfloor, \lfloor y^{i,j_2}\rfloor)\} + 2\sqrt{2}\epsilon L \\
&\leq \sqrt{2}\epsilon L + \sqrt{2}\epsilon L + 2\sqrt{2}\epsilon L \\
&= 4\sqrt{2}\epsilon L
\end{aligned} \quad (10)$$

Proceeding by induction, we know that

$$\xi^i \leq 2\sqrt{2}(|J|-1)\epsilon L, \forall i \in I$$

since an inside sensor can have at most $|J|$ adjacent outside sensors.

Now, suppose that $i_1, i_2 \in I, i_1 \neq i_2$. Then,

$$\begin{aligned}
\xi^{\{i_1,i_2\}}(X^{\{i_1,i_2\}}, Y^{\{i_1,i_2\}}) &= t^{\{i_1,i_2\}}(X^{\{i_1,i_2\}}, Y^{\{i_1,i_2\}}) - T^{\{i_1,i_2\}}(X^{\{i_1,i_2\}}, Y^{\{i_1,i_2\}}) \\
&= t^{i_1}(x^{i_1}, y^{i_1}) + t^{i_2}(x^{i_2}, y^{i_2}) - T^{i_1}(X^{i_1}, Y^{i_1}) - T^{i_2}(X^{i_2}, Y^{i_2}) \\
&\leq t^{i_1}(x^{i_1}, y^{i_1}) + t^{i_2}(x^{i_2}, y^{i_2}) - T^{i_1}(\lfloor x^{i_1}\rfloor, \lfloor y^{i_1}\rfloor) - T^{i_2}(\lfloor x^{i_2}\rfloor, \lfloor y^{i_2}\rfloor) \\
&\leq t^{i_1}(\lceil x^{i_1}\rceil, \lceil y^{i_1}\rceil) + t^{i_2}(\lceil x^{i_2}\rceil, \lceil y^{i_2}\rceil) - T^{i_1}(\lfloor x^{i_1}\rfloor, \lfloor y^{i_1}\rfloor) - T^{i_2}(\lfloor x^{i_2}\rfloor, \lfloor y^{i_2}\rfloor) \\
&= \{t^{i_1}(\lfloor x^{i_1}\rfloor, \lfloor y^{i_1}\rfloor) - T^{i_1}(\lfloor x^{i_1}\rfloor, \lfloor y^{i_1}\rfloor)\} + \{t^{i_2}(\lfloor x^{i_2}\rfloor, \lfloor y^{i_2}\rfloor) - T^{i_2}(\lfloor x^{i_2}\rfloor, \lfloor y^{i_2}\rfloor)\} + \\
&\quad + \{t^{i_1}(\lceil x^{i_1}\rceil, \lceil y^{i_1}\rceil) - t^{i_1}(\lfloor x^{i_1}\rfloor, \lfloor y^{i_1}\rfloor)\} + \{t^{i_2}(\lceil x^{i_2}\rceil, \lceil y^{i_2}\rceil) - t^{i_2}(\lfloor x^{i_2}\rfloor, \lfloor y^{i_2}\rfloor)\} \\
&\leq 2\sqrt{2}(|J|-1)\epsilon L + 2\sqrt{2}(|J|-1)\epsilon L + 2\sqrt{2}\epsilon L \\
&\leq 4\sqrt{2}(|J|)\epsilon L
\end{aligned} \quad (11)$$

We can also bound the error in the case of three inside sensors and all of their adjacent outside sensors:

$$\begin{aligned}
\xi^{\{i_1,i_2,i_3\}}(X^{\{i_1,i_2,i_3\}}, Y^{\{i_1,i_2,i_3\}}) &= t^{\{i_1,i_2,i_3\}}(X^{\{i_1,i_2,i_3\}}, Y^{\{i_1,i_2,i_3\}}) - T^{\{i_1,i_2,i_3\}}(X^{\{i_1,i_2,i_3\}}, Y^{\{i_1,i_2,i_3\}}) \\
&= t^{\{i_1,i_2\}}(x^{\{i_1,i_2\}}, y^{\{i_1,i_2\}}) + t^{i_3}(x^{i_3}, y^{i_3}) - T^{\{i_1,i_2\}}(X^{\{i_1,i_2\}}, Y^{\{i_1,i_2\}}) - T^{i_3}(X^{i_3}, Y^{i_3}) \\
&\leq t^{\{i_1,i_2\}}(x^{\{i_1,i_2\}}, y^{\{i_1,i_2\}}) + t^{i_3}(x^{i_3}, y^{i_3}) - T^{\{i_1,i_2\}}(\lfloor x^{\{i_1,i_2\}}\rfloor, \lfloor y^{\{i_1,i_2\}}\rfloor) - T^{i_3}(\lfloor x^{i_3}\rfloor, \lfloor y^{i_3}\rfloor) \\
&\leq t^{\{i_1,i_2\}}(\lceil x^{\{i_1,i_2\}}\rceil, \lceil y^{\{i_1,i_2\}}\rceil) + t^{i_3}(\lceil x^{i_3}\rceil, \lceil y^{i_3}\rceil) - T^{\{i_1,i_2\}}(\lfloor x^{\{i_1,i_2\}}\rfloor, \lfloor y^{\{i_1,i_2\}}\rfloor) - T^{i_3}(\lfloor x^{i_3}\rfloor, \lfloor y^{i_3}\rfloor) \\
&= \{t^{\{i_1,i_2\}}(\lfloor x^{\{i_1,i_2\}}\rfloor, \lfloor y^{\{i_1,i_2\}}\rfloor) - T^{\{i_1,i_2\}}(\lfloor x^{\{i_1,i_2\}}\rfloor, \lfloor y^{\{i_1,i_2\}}\rfloor)\} + \{t^{i_3}(\lfloor x^{i_3}\rfloor, \lfloor y^{i_3}\rfloor) - T^{i_3}(\lfloor x^{i_3}\rfloor, \lfloor y^{i_3}\rfloor)\} + \\
&\quad + \{t^{\{i_1,i_2\}}(\lceil x^{\{i_1,i_2\}}\rceil, \lceil y^{\{i_1,i_2\}}\rceil) - t^{\{i_1,i_2\}}(\lfloor x^{\{i_1,i_2\}}\rfloor, \lfloor y^{\{i_1,i_2\}}\rfloor)\} + \{t^{i_3}(\lceil x^{i_3}\rceil, \lceil y^{i_3}\rceil) - t^{i_3}(\lfloor x^{i_3}\rfloor, \lfloor y^{i_3}\rfloor)\} \\
&\leq 4\sqrt{2}|J|\epsilon L + 2\sqrt{2}(|J|-1)\epsilon L + 2\sqrt{2}\epsilon L \\
&\leq 6\sqrt{2}(|J|)\epsilon L
\end{aligned} \quad (12)$$

Proceeding by induction, we know that $\xi^I$ the error of the discrete approximation for the entire set of internal sensors $I$ and all of their adjacent outside sensors is bounded by

$$\xi^I \leq 2\sqrt{2}|I||J|\epsilon L$$

Therefore,

$$\lim_{\epsilon \to 0} \xi^I = t^I(X^I, Y^I) - T^I(X^I, Y^I)$$
$$\leq \lim_{\epsilon \to 0} 2\sqrt{2}|I||J|\epsilon \qquad (13)$$
$$= 0$$

Hence, as $\epsilon \to 0$ the discrete approximation $T^I(X^I, Y^I)$ converges to the true continuous optimal detection value $t^I(X^I, Y^I)$.

# 7 The Case of an Adaptive Adversary

So far, our model assumed a fixed flow of dangerous material on each pathway, and we have presented a method that would allow law enforcement officials to use current information on attacker behavior to maximize the amount of captured illegal or dangerous contraband. However, we can think of attackers as intelligent adversaries who would adjust their strategy once they observe the changes in site security. Therefore, the goal of a defensive strategy could be to make sure that no path leading into the site has a violation detection rate that is unreasonably low. For example, suppose we have an adaptive adversary who recognizes how much of a resource we use for sensors on each node and then chooses the path that minimizes the probability of detection. To defend against such an adversary we might seek to assign sensor resources so as to maximize the minimum detection rate on any path. Hence we face the following optimization challenge:

$$\max_{\mathbf{x},\mathbf{y}} \min_{\substack{i \in I \\ j \in N(i)}} \left\{ D_j^y(y_j) + D_i^x(x_i)(1 - D_j^y(y_j)) \right\}$$
$$\text{s.t.} \sum_{i \in I} x_i \leq X$$
$$\sum_{j \in J} y_j \leq Y \qquad (14)$$
$$x_i \geq 0, \forall i \in I$$
$$y_j \geq 0, \forall j \in J$$

In order to solve this problem we can use a similar approach to the one discussed in the previous section.

**Steps 1 & 2:** These are identical to their counterparts described in Section 4, with $F_j = 1$ for every outside sensor $j$.

**Step 3:**
Once again, we denote by $T^{i,j}$ the resulting table of values generated at Step 2. More specifically, we denote with $T^{i,j}(X_i, Y_j)$ the optimal detection value that can be achieved by investing $(X_i, Y_j) \in \mathcal{X} \times \mathcal{Y}$ resources of respectively, inner and outer resources. Then we consider the case of two outside sensors $j_1, j_2$ that are backed up by inner sensor $i$. We can merge $T^{i,j_1}$ and $T^{i,j_2}$ into a single solution according to:

$$T^{i,\{j_1,j_2\}}\left(X^{i,\{j_1,j_2\}}, Y^{i,\{j_1,j_2\}}\right) = \begin{cases} \max_{y^{i,j_1}, y^{i,j_2}} \min \left\{ T^{i,j_1}\left(X^{i,\{j_1,j_2\}}, y^{i,j_1}\right), T^{i,j_2}\left(X^{i,\{j_1,j_2\}}, y^{i,j_2}\right) \right\} \\ \text{s.t.} \\ y^{i,j_1} + y^{i,j_2} \leq Y^{i,\{j_1,j_2\}} \\ y^{i,j_1} \in \mathcal{Y} \\ y^{i,j_2} \in \mathcal{Y} \end{cases} \qquad (15)$$

where $T^{i,\{j_1,j_2\}}(X^{i,\{j_1,j_2\}}, Y^{i,\{j_1,j_2\}})$ is the optimal detection value for the combined system.

Again, if we proceed by induction, we can generate an optimal value table for a problem instance that involves one inner sensor $i$, and an arbitrary number of outside sensors. We denote such a table by $T^i$.

**Step 4:**

Consider two matrices $T^{i_1}$ and $T^{i_2}$ that correspond to respectively inner sensors $i_1$ and $i_2$ with all of their adjacent outside sensors. We could again merge the two solutions into a single global solution according to the following rule:

$$T^{\{i_1,i_2\}}(X^{\{i_1,i_2\}}, Y^{\{i_1,i_2\}}) = \begin{cases} \max_{x^{i_1}, x^{i_2}, y^{i_1}, y^{i_2}} \min\left\{ T^{i_1}(x^{i_1}, y^{i_1}), T^{i_2}(x^{i_2}, y^{i_2}) \right\} \\ \text{s.t.} \\ x^{i_1} + x^{i_2} \leq X^{\{i_1,i_2\}} \\ y^{i_1} + y^{i_2} \leq Y^{\{i_1,i_2\}} \\ x^{i_1}, x^{i_2} \in \mathcal{X} \\ y^{i_1}, y^{i_2} \in \mathcal{Y} \end{cases} \tag{16}$$

where $T^{\{i_1,i_2\}}(X^{\{i_1,i_2\}}, Y^{\{i_1,i_2\}})$ denotes the optimal value for the combined system that employs internal sensors $i_1$ and $i_2$, and all of their outside neighbors.

Once again, if we have a third branch consisting of inside sensor $i_3$ and its outside neighbours, then we can use $T^{\{i_1,i_2\}}$ and $T^{i_3}$ as inputs to equation (16) and find the table of optimal values $T^{\{i_1,i_2,i_3\}}$ for the combined system consisting of inside sensors $i_1, i_2, i_3$, and all of their adjacent outside sensors. Proceeding by induction, we know that even if we have an adaptive adversary, we can solve problems involving an arbitrary number of interior and exterior sensors, as well as sensor detection curves specified by concave increasing piecewise linear functions.

# 8 Computational Results

In this section we present computational results for the methods developed above. The experiments were performed on an AMD Phenom X4 9550 workstation with 6GB of DDR2 RAM. We consider two different system configurations and for each of them we provide plots of the objective function value for both the original and adaptive adversary models.

*Example* 8.1. In this example we consider an inner layer consisting of four sensors one with three adjacent outside sensors (indices 1, 2, 3), a second with two adjacent outside sensors (indices 4, 5), a third with two adjacent outside sensors (indices 6, 7), and a fourth with two adjacent outside sensors (indices 8, 9). For each inside sensor $i \in I$, we specify $D_i^x(x) = \min\{0.2x, 0.4 + 0.1x\}$. Further, for the first outside sensor we use $D_{j_1}^y(y) = \min\{0.3y, 0.3 + 0.1y, 0.5 + 0.05y\}$, and for outside sensors of index $j = 2, 3, \ldots, 9$, we use $D_j^y(y) = \min\{0.3y, 0.3 + 0.1y\}$. In addition, all outside sensors have exactly 1 unit of incoming flow. Figure 11 gives the solution maximizing the expected amount of captured contraband for a range of interior and exterior budgets, i.e., the solution to the first problem. The solution matrix includes 10,302 distinct points and the computation took 117 seconds.

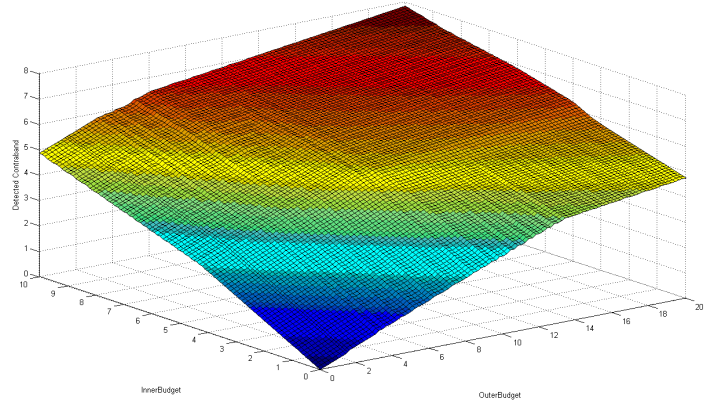

**Figure 11:** Solution maximizing the expected amount of captured contraband for a range of interior and exterior budgets for Example 8.1.

We can also calculate the adaptive adversary solution maximizing the minimum probability of capturing contraband along all paths for a range of interior and exterior budgets. The solution matrix shown in Figure 12 includes 40,401 distinct points and the computation took 3,102 seconds (52 minutes).

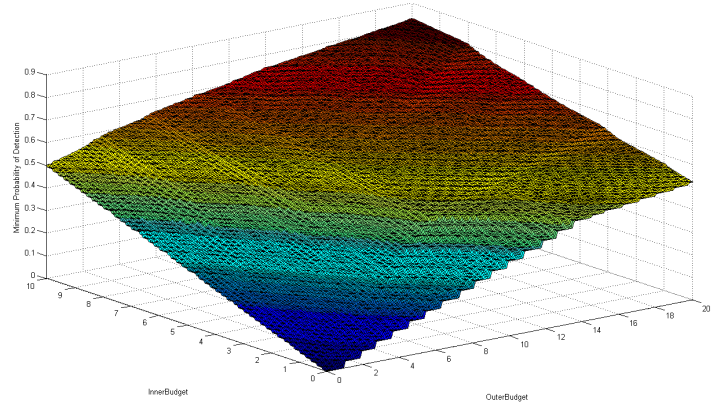

**Figure 12:** Adaptive adversary solution maximizing the minimum probability of capturing contraband along all paths for a range of interior and exterior budgets for Example 8.1.

*Example* 8.2. In this example, we modify Example 8.1 so that all the outside sensors have exactly 1 unit of incoming flow except for outside sensors 1 and 9 which have 10 units of incoming flow. Figure 13 shows the optimal objective values of the maximized amount of captured contraband for a range of interior and exterior budgets. The solution table includes 10,302 distinct points, and the computation took 119 seconds.

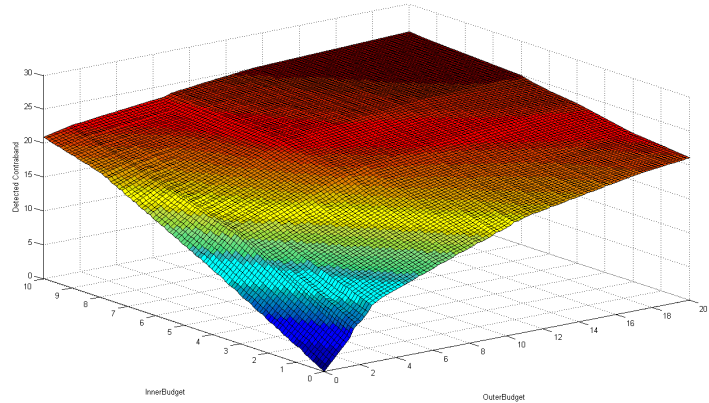

**Figure 13:** Solution maximizing the expected amount of captured contraband for a range of interior and exterior budgets for Example 2.

In this example we only changed the flow values of Example 8.1. For this reason, we do not need to compute a new adaptive adversary solution, as it would be identical to the one for Example 8.1. Naturally, this illustrates the robustness of the adaptive adverdsary formulation compared to its original counterpart.

## 9 Closing Remarks

We have considered the problem of determining the optimal resource allocation for layered security. A computational method for the maximization of captured contraband and an adaptive adversary approach for the maximization of the worst case probability of detection have been developed. Both methods are computationally tractable and can be applied to non-trivial practical problems.

We have a great deal more that we can do in the future. One thing is to consider both legal and illegal flow, which we also refer to as respectively good and bad units. Hence, in addition to detecting bad units we could consider false positive decisions for each sensor and adopt a risk-averse optimization approach [3]. Another possible direction would be attempting to write the problem as a large game and use approximation methods similar to the ones developed by Grigoriadis and Khachian [10]. Alternatively, we could look into interdiction on planar graphs methods similar to the ones developed by Zenklusen [18, 19].

Still another approach is to follow the applications of Stackelberg games that have been used in pioneering defensive approaches at the nation's airports, ports, and in applications by the Federal Air Marshals Service. US Coast Guard, etc. (see [11, 15]). In a Stackelberg game between an attacker and a defender, the defender (security) acts first. The attacker can observe the defender's strategy and choose the most beneficial point of attack. The challenge is to introduce some randomness in the defender's strategy to increase the uncertainty on the part of the attacker. Bayesian Stackelberg games do exactly that. Layered defense makes this into a new kind of Stackelberg game to analyze, one with two rounds, one involving the outer layer and one involving the inner layer based on results at the outer layer. We can look both at nonrandomized and randomized strategies for the defender.

There are many other directions in which this work could go. Even with our current model, we have not yet developed practical methods to handle more than two layers of defense. There are also many variations on our model that could be quite interesting. For example, we could consider a fixed resource limit that the defender could allocate between inner and outer layers. Then, we could allow adaptive redistribution of resources across layers and across time (see [2, 4]).